\documentclass{article}
\usepackage{graphicx}
\usepackage{color}   
\usepackage{bm}
\usepackage[margin=2cm]{geometry}
\newcommand{\rem}[1]{}

\title{Reconfigurable assembly of nematic colloids commanded by photoactivated surface patterns}
\author{Sergi Hern\`{a}ndez-Navarro$^{1}$,\and Pietro Tierno$^{2,3}$,\and Joan Anton Farrera$^{4}$,\and Jordi Ign\'{e}s-Mullol$^{1,3}$,\and Francesc Sagu\'{e}s$^{1,3}$}
\begin{document}
\maketitle
\begin{enumerate}
 \item Department of Physical Chemistry, University of Barcelona, Mart\'{\i} i Franqu\`es 1, 08028 Barcelona, Spain.
 \item Department of Structure and Constituents of Matter, University of Barcelona, Avinguda Diagonal 647, 08028 Barcelona, Spain.
 \item Institute of Nanoscience and Nanotechnology, IN$^2$UB, University de Barcelona, Barcelona, Spain.
 \item Department of Organic Chemistry, University of Barcelona, Mart\'{\i} i Franqu\`es 1, 08028 Barcelona, Spain.
\end{enumerate}

{\bf Abstract}. Different phoretic driving mechanisms have been proposed for the transport of solid or liquid microscopic inclusions in integrated chemical processes. However, the ability to reversibly address both the flow path, rate, and local reactant concentration has not yet been realized. Here we show that a substrate chemically modified with photosensitive self-assembled monolayers allows to directly command the assembly and transport of large ensembles of micron-size particles and drops dispersed in a thin layer of anisotropic fluid. Our strategy separates particle driving, realized via AC electrophoresis, and steering, achieved by elastic modulation of the host nematic fluid. Inclusions respond individually or in collective modes following arbitrary reconfigurable paths imprinted via UV/blue illumination.  Relying solely on generic material properties, the proposed procedure is versatile enough for the development of applications involving either inanimate or living materials.

\clearpage
%
%
In the emerging field of microfluidics, analytic operations are usually performed by pumping chemicals through channels that are typically hundreds of micrometers wide \cite{kim1995,squires2005}. While strategies to steer fluids are well established through the use of photolithographic networks, further versatility of integrated lab-on-a-chip devices demands the ability to transport femtosized volumes of chemical cargo \cite{song2006} using approaches that allow to externally command their collective dynamics and assembly without relying on permanent geometrical constraints.

Colloidal suspensions, formed by microsized inclusions embedded in a carrier fluid are a natural starting point for such a strategy. The self-assembly of colloidal entities can be tuned by controlling inclusion size\cite{yethiraj2004} and shape\cite{glotzer2007}, surface chemistry\cite{jiang2010,wang2012}, or by suspending the colloids in a structured liquid\cite{poulin1997}. Recently, the focus has shifted into managing dynamic self-assembly\cite{whitesides2002}, seeking, for instance, the emergence of collective behavior \cite{duan2013}, which may result in novel materials with enhanced functionalities\cite{zheludev2012}.

Different strategies exist to drive individual colloids dispersed in a liquid\cite{grier2003,paxton2004,paxton2006,howse2007,wang2009,buttinoni2013,wang2013,gao2014}, and to command their assembly\cite{ibele2009,kagan2011,duan2013}, although the control and large-scale addressability of collections of motile inclusions has proven challenging to realize. Optical traps are often used to achieve direct control over the placement of colloids, with holographic tweezers allowing to extend such control to a few hundreds of inclusions\cite{grier2003}, although the technique is limited by the field of view of the optical system. Using a completely different approach, large-scale motion of independent microscopic inclusions has been demonstrated on chemically and physically activated colloids.\cite{paxton2004,paxton2006,howse2007,wang2009,buttinoni2013,wang2013,gao2014} Reconfigurable assembly of colloidal clusters has been achieved through the chemically adjustable competition between electrophoretic attraction and electroosmotic repulsion effects \cite{ibele2009,duan2013}. Recently, these ideas have been combined to realize the assembly and transport of small clusters of phoretic colloids driven by osmotic pressure imbalances\cite{palacci2013}. These results have opened new perspectives for reconfigurable self-assembly, enabling the massive transport of inclusions\cite{bricard2013}, although the precise and selective control over swarms of colloids has not yet been demonstrated.

Of particular interest is the use of alternating current (AC) electric fields to transport colloidal inclusions. Since AC fields avoid the ion migration mechanisms caused by direct current driving, they would be preferred for the transport of water-based liquid inclusions or living cargo. While individual metallo-dielectric Janus particles can be driven by this method in aqueous media\cite{gangwal2008}, nematic liquid crystal (NLC) host allowed to use AC electrophoresis to transport homogeneous solid\cite{lavrentovich2010,lavrentovich2014} or liquid inclusions\cite{hernandez2011}. A NLC is an oil featuring long-range orientational order (the director field) of its constituent molecules, but can also be realized in concentrated aqueous suspensions of amphiphiles (lyotropic liquid crystals)\cite{oswald}. One of the most intriguing characteristics of these viscoelastic materials is their interaction with non-flat surfaces, in particular colloidal inclusions, that leads to the proliferation of topological point and line defects. This feature has been exploited in recent years to develop colloid-based self-assembling materials\cite{musevic2006,wood2011} and can be used as a positioning mechanism for embedded colloids\cite{martinez2011}.

We integrate all the above possibilities in a novel electro-optical technique that permits to remotely address the reversible assembly and collective transport of micron-size colloidal inclusions of arbitrary shape and composition, both solid and liquid, dispersed in a thin NLC film. We use liquid crystal-enabled electrophoresis (LCEEP)\cite{lavrentovich2014} to propel the colloidal inclusions embedded in a NLC cell, and a photosensitive anchoring layer to modify the local director field, which, in turn, sets the direction of colloidal motion. We assemble or disassemble swarms, and control their placement and motion over arbitrary paths on a surface.

\begin{figure}
\includegraphics[width=\textwidth,keepaspectratio]{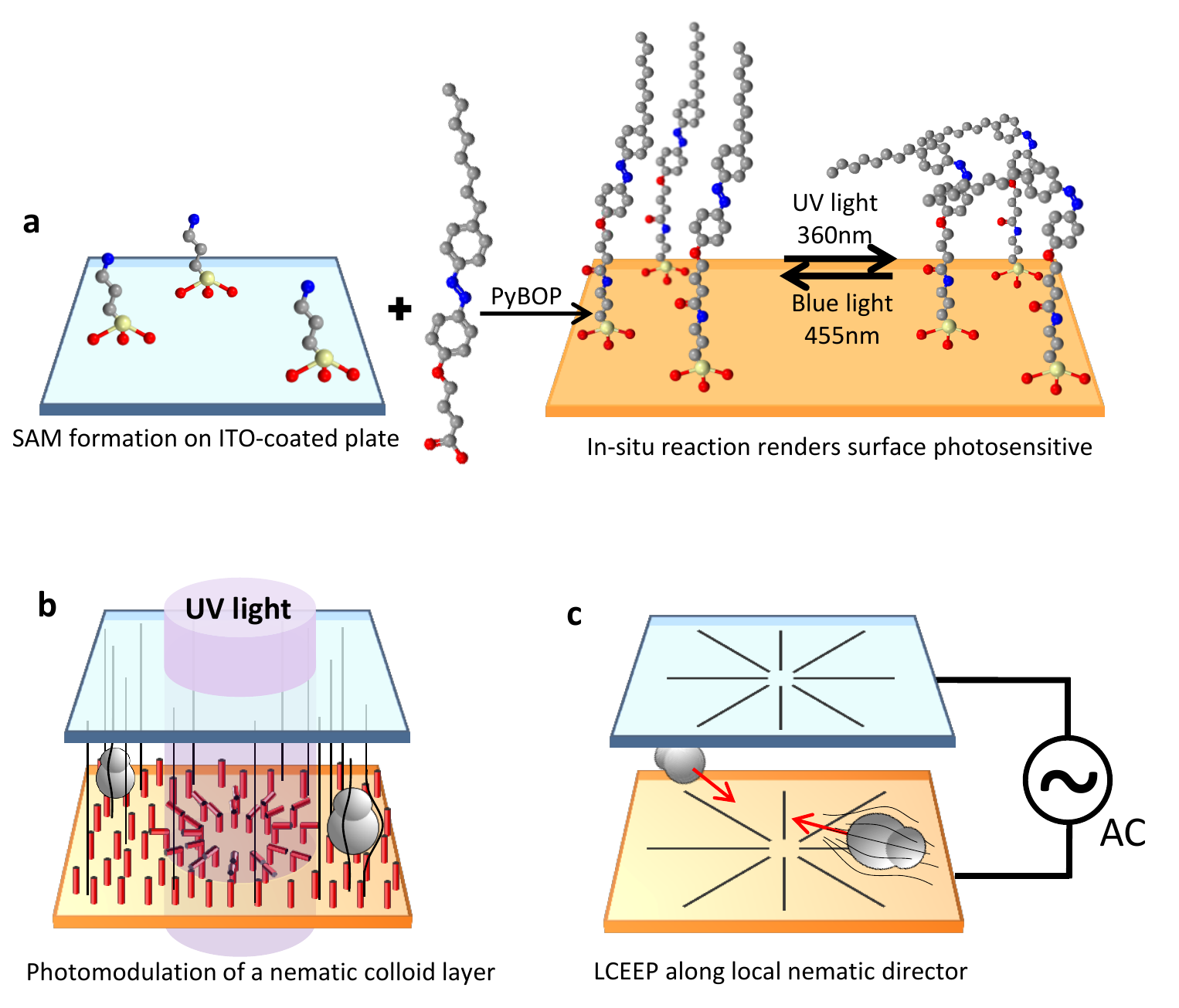}%
\caption{
\label{fig:experimental}%
Schematics showing the preparation of photo-addressable nematic colloids.
\textbf{a} Representation of the two-step surface functionalization to prepare a photosensitive ITO coated glass surface. The grafted alkyl-azobenzene chains can be reversibly switched between the \emph{cis} and \emph{trans} isomers. \textbf{b} Colloids are dispersed in a nematic liquid crystal confined between a photosensitive and a non photosensitive plate. This allows to reversibly light-scribe patterns of in-plane alignment (here, a circular spot characterized by a radial pattern). The embedded anisometric particles are aligned by the local nematic director (black lines). Upon application of an AC electric field (\textbf{c}) planar regions expand, and the particles are electrophoretically driven along the local director.
}
\end{figure}

Our system consists of a NLC film inserted between parallel glass plates coated with a transparent electrode, which allows application of electric fields perpendicular to the plates.
One of the plates is functionalized with an azosilane photosensitive self-assembled monolayer that allows to alternate between perpendicular (homeotropic) or planar (tangential) boundary conditions (anchoring) of the contacting NLC. The counter plate is treated with a spin-coated polyimide compound to achieve strong and permanent homeotropic anchoring. Without external influences, these boundary conditions lead to uniform homeotropic anchoring of the NLC. By irradiating with UV (365 nm) light from an incoherent source we force the azosilane monolayer to adopt the \emph{cis} (planar) configuration, which can be easily reverted to \emph{trans} (homeotropic) when using blue (455 nm) illumination (Fig. \ref{fig:experimental} and Fig. S1).
We employ a NLC with negative dielectric anisotropy, which aligns parallel to the glass plates upon application of the electric field. This leads to degenerate planar alignment conditions, since all in-plane orientations for the director field are energetically equivalent, thus enabling the local addressability of the NLC director.

As colloidal inclusions we use pear-shaped microparticles made of polystyrene, a material that promotes planar orientation of the NLC on the particle surface (Methods Summary and Section S1). The chosen particle shape guarantees a dipolar component in the distortion of the local director field, a requirement for LCEEP to be an efficient propulsion mechanism. Earlier evidences of this mechanism have employed spherical solid\cite{lavrentovich2010} or liquid\cite{hernandez2011} inclusions, relying on surface functionalization to achieve a homeotropic anchoring on the particle surface, which leads to a dipolar arrangement of the NLC director around the inclusion. However, homeotropic anchoring has the disadvantage that it can degenerate into a quadrupolar director configuration for which LCEEP propulsion is not effective. Our choice thus ensures that all particles in a large ensemble are similarly propelled.

\begin{figure}
\includegraphics[width=\textwidth,keepaspectratio]{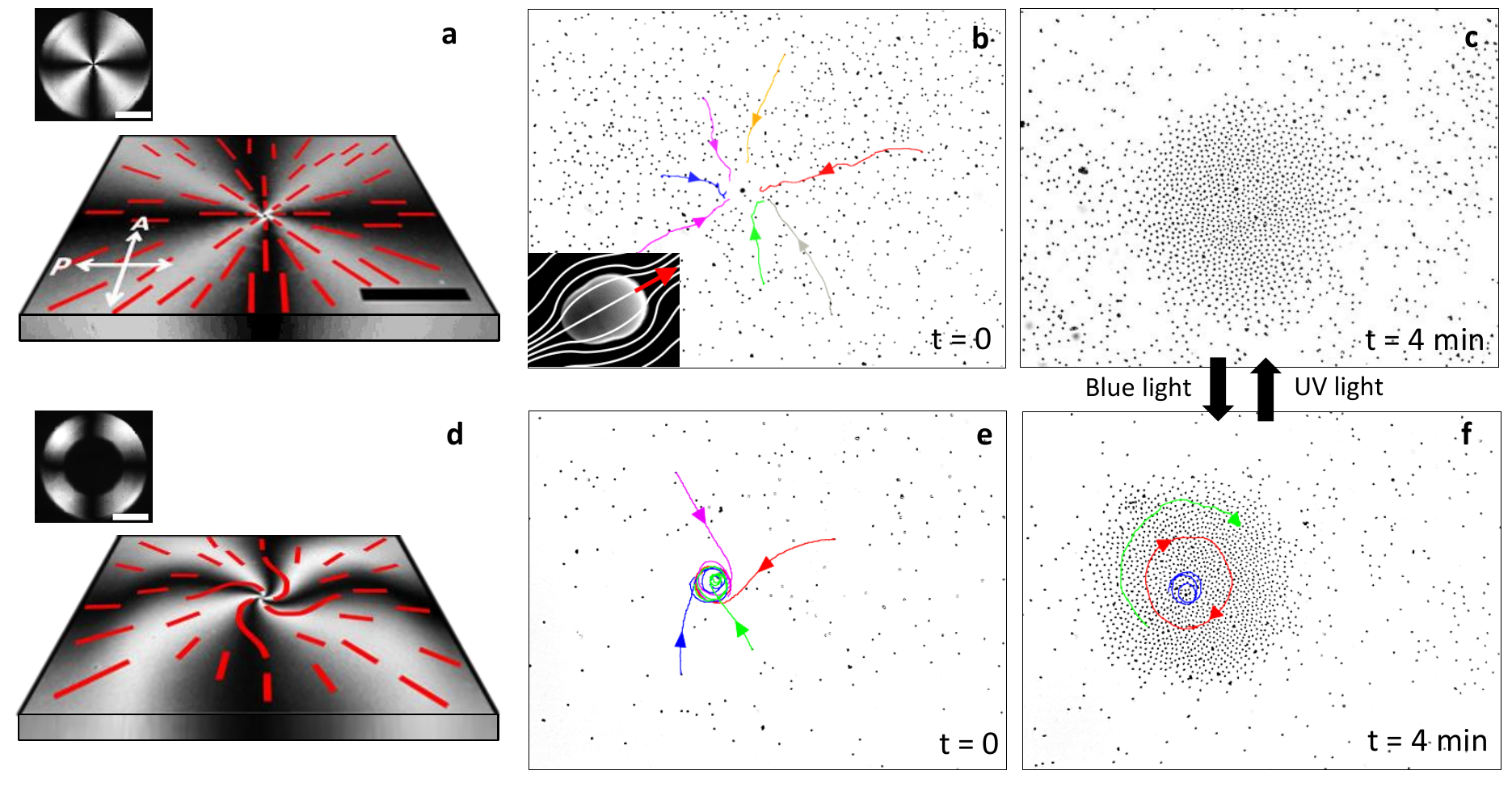}%
\caption{
\label{figure1}
Formation of nematic colloidal clusters.
Top (\textbf{a}-\textbf{c}) [bottom, (\textbf{d}-\textbf{f})] row of images illustrating the formation of a colloidal aster (vortex).
(\textbf{a}) and (\textbf{d}) are images between crossed polarizers of the imprinted NLC texture leading to
a cross (\textbf{a}) or a spiral (\textbf{d}) attracting pattern. Dashed red lines represent the NLC director orientation.
The insets show the planar photoaligned circle (inset \textbf{a}) or corona (inset \textbf{d}) prior to application of the electric field.
Sequence \textbf{b}-\textbf{c} (\textbf{e}-\textbf{f}) shows the formation of an aster (vortex) after application of an electric field with frequency 10 Hz and amplitude 0.87 V/$\mu$m. The trajectories followed by several particles are superimposed to the images to illustrate the cluster formation mechanism.
Inset in (\textbf{b}) shows a S.E.M. image of a single pear-shaped particle ($3 \mu m\times 4 \, \mu m$), with NLC field lines and the direction of motion.
The colloidal aster in (\textbf{c}) and vortex in (\textbf{f}) can be interconverted by suitable irradiation protocols, as explained in the text. The scale bars are $200 \, \mu$m for all images and $500\, \mu$m for the two insets (\textbf{a} and \textbf{d}).
}
\end{figure}

 In the absence of irradiation or electric field, particles will be oriented perpendicularly to the cell plates, following the uniform NLC director. Shinning the cell for a few seconds with a spot of UV light (Fig. S2) forces the NLC in contact with the azosylane-treated surface to transit to a planar configuration. As shown in Fig. \ref{figure1}a and Fig. S3, this configuration conforms to the applied irradiation (Fig. S6) by locally adopting a splay (radially-spread) texture organized around a central defect\cite{oswald}. Application of an external AC field forces the bulk NLC to adopt the splay configuration that now extends for several millimeters, well beyond the size of the irradiated spot, thanks to the anchoring degeneracy. The induced configuration is stable for days under AC field, well past the half-life for thermal relaxation of the azosilane film, which is about 30 minutes (Fig. S7). The region with radial alignment will be the basin of attraction for dispersed particles, which tumble instantaneously following the NLC director so that their long axis lays, on average, parallel to the cell plates (Fig. S3). Simultaneously, LCEEP sets the particle into motion at a constant speed. For a given particle size, this velocity can be tuned with the intensity and the frequency of the AC field (Fig. S8). This feature allows, for instance, to decouple NLC realignment and particle motion, since their speed is negligible for field frequencies above 50 Hz, while NLC realignment can be achieved even for frequencies in the kHz range and above\cite{oswald}. All particles move following the local NLC director, with roughly half of them being attracted by the photoinduced radial defect and half of them being repelled by it. This is a consequence of the random tumbling of the particle dipole from their initial orientation perpendicular to the planar alignment induced by the AC field (Figs. S4 and S9). As shown in Fig. \ref{figure1}b-c (Movie S1), particles follow the NLC field lines and assemble into a growing aster-like configuration where they jam and become arrested.

We can easily switch the assembling mode between an arrested aster and a rotating vortex cluster by taking advantage of the elastic properties of the dispersing NLC, and the fact the particles are slave to the NLC director field. This is achieved by \emph{erasing} the central region of an imprinted UV area with a smaller spot of blue light, which forces the \emph{cis} to \emph{trans} isomerisation leading to homeotropic NLC alignment. As shown in Fig. \ref{figure1}d and inset, upon application of the AC field the planar alignment in the resulting circular corona is extended both outwards and inwards. The region surrounding the inner defect now features degenerate planar anchoring conditions, so it relaxes from the pure splay texture to the less energy demanding bend (rotational) texture\cite{elastic_constants}. As a consequence, particles follow spiral trajectories, and assemble into a rotating vortex, preceding around the central defect with a constant linear velocity, as seen in Fig. \ref{figure1}e,f (Movie S1 and Fig. S10). The vortex cluster can be transformed back into an arrested aster by irradiating the region with a UV light spot. Both assembly modes can thus be reversibly interconverted in real time via photoelastic control (Movie S1). The observed dynamic phenomena can be captured by a simple phenomenologial model to reproduce the coupling between electrophoretic driving and the underlying photoactivated surface patterns (see Supplementary Text, Fig. S11, and Movie S2).

\begin{figure}
\includegraphics[width=\textwidth,keepaspectratio]{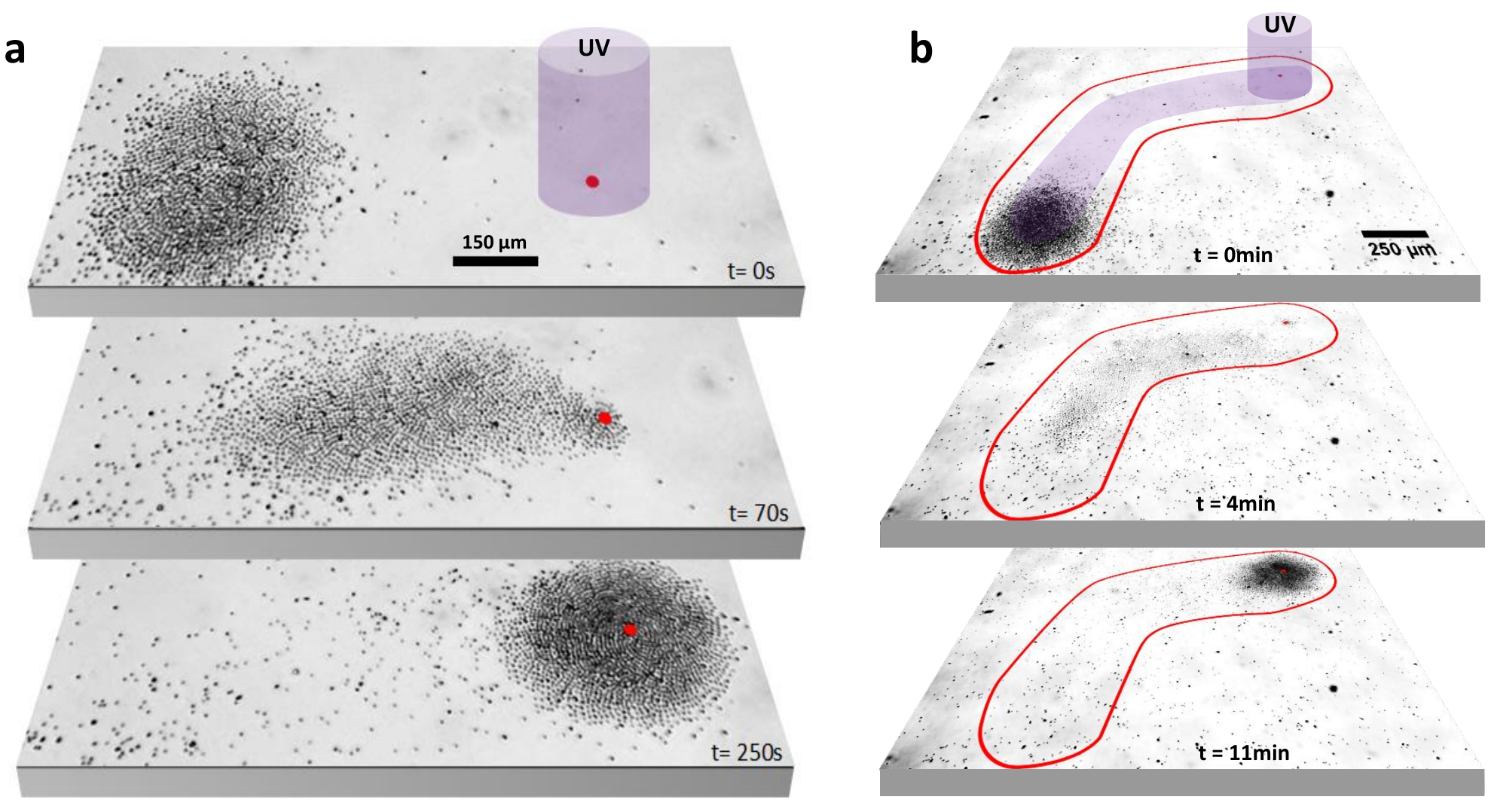}%
\caption{\label{filling}
Cluster addressability.
Image sequence of a particle swarm traveling across the LC cell due to in situ reconfiguration of the NLC field. In (\textbf{a}), The photoaligned spot initially centered with the cluster is moved 600 $\mu$m to the right in a straight, as indicated by the red spot. In (\textbf{b}) a longer hybrid track combining curved and straight segments is demonstrated. The contour of the track, only visible between crossed polarizers, is outlined in red. The applied sinusoidal electric field has an amplitude of 0.74 V/$\mu$m and a frequency of 10 Hz.
 }
\end{figure}

The reversibility and quick response of the photoalignment layer enables straightforward cluster addressability. A preformed aggregate of arbitrary size, either aster or vortex-like, can be relocated to a pre-designed place anywhere within the experimental cell with minimum dismantlement of the cluster structure by changing the location of the UV spot. An example of this process is shown in Fig. \ref{filling}a (also Movie S3 and Fig. S4). After blocking the LCEEP mechanism by increasing the field frequency above 50 Hz, the center of attraction is translated 600 $\mu$m to the right. Once LCEEP is reactivated, the swarm of particles moves towards the new position developing a leading edge around which the particles assemble. Alternatively, by the same principle one can imprint predesigned arbitrary paths connecting distant locations inside the cell or draw circuits with complex topologies as a simple way to accumulate colloidal swarms in the irradiated area and further entrain them collectively as shown in Fig. \ref{filling}b (Movie S3).

\begin{figure}
\includegraphics[width=\textwidth,keepaspectratio]{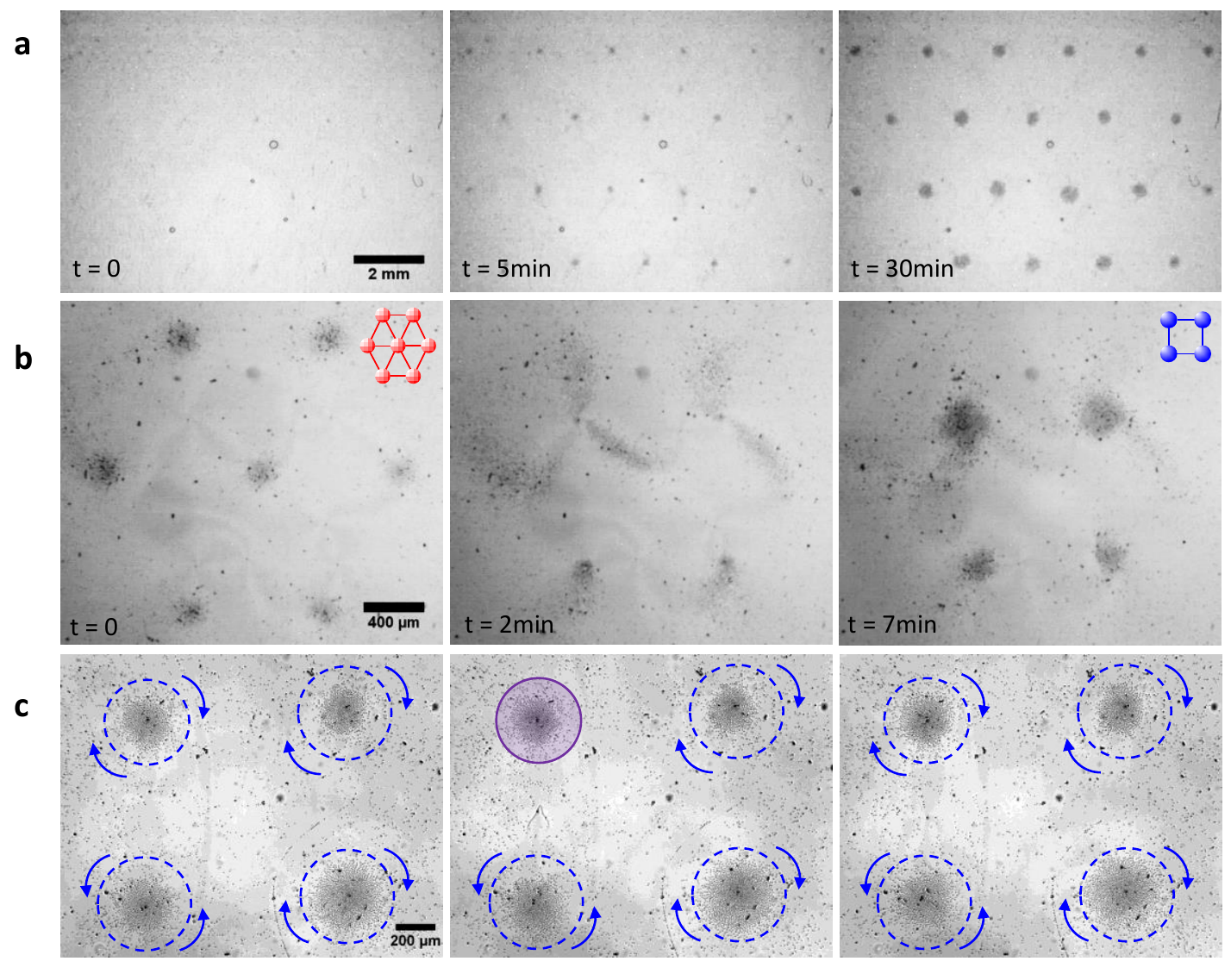}%
\caption{
\label{fig:hex_lattice}
Reconfigurable cluster lattices.
 \textbf{a} Image sequence of the formation of a triangular lattice of particle clusters. A sinusoidal electric field of 0.74 V/$\mu$m and 10 Hz is used to drive the particles. \textbf{b} Image sequence of the transformation from a triangular to a square lattice. The original distribution of attractor spots is erased with blue light (455 nm), and a square lattice is subsequently imprinted using UV light (365 nm). The cluster rearrangement occurs when a sinusoidal electric field of 0.78 V/$\mu$m and 10 Hz is applied. Swarms of particles subsequently form and redistribute into the reconfigured lattice. \textbf{c} The dynamic state of clusters on a lattice can be individually addressed. While all the clusters are in a rotating dynamics, the top left cluster is first arrested and then restarted.}
\end{figure}

Our strategy to command colloidal aggregation can be readily implemented for the direct or indirect transport of chemical cargo in a channel-free microfluidic environment.  For instance, one can build photoactivated lattices of clusters with arbitrary symmetry on an extended surface by imprinting the corresponding distribution of irradiated spots. As an example, in Fig. \ref{fig:hex_lattice}a (Movie S4) a triangular lattice of UV spots is imprinted onto the photosensitive surface. Upon application of the electric field, these spots compete as attractors for the particles dispersed in the NLC. Rearrangement of the entire lattice, and thus redistribution of chemical cargo, is readily achieved by conveniently reshaping the pattern of projected light spots. We demonstrate this addressability in Fig. \ref{fig:hex_lattice}b (Movie S4), where the initially triangular lattice is transformed into a square one. Swarms of particles redistribute in the NLC cell according to the new director landscape. Once the electric field is switched off, the instantaneous pattern of clusters is preserved for an extended amount of time. For the used NLC, with a dynamic viscosity $\eta\simeq$ 0.1 Pa.s, we estimate a self-diffusion coefficient $D \simeq 10^{-3}\mu$m$^2$s$^{-1}$. Thus, it would take several hours for particles of the size used here to diffuse a few microns, and the spontaneous disaggregation of a cluster could take months.

\begin{figure}
\includegraphics[width=\textwidth,keepaspectratio]{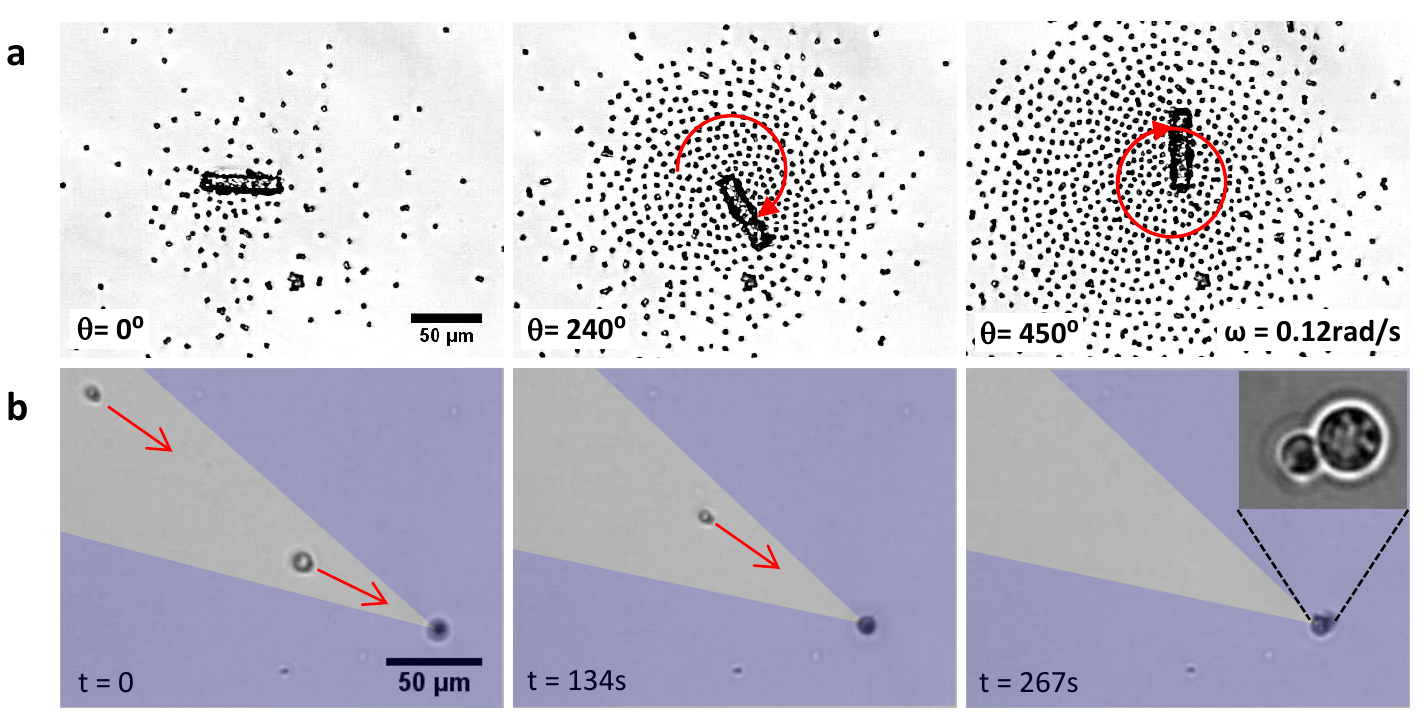}%
\caption{
\label{fig:devices}
Microfluidic applications.
 \textbf{a} A glass microrod (50 $\mu$m long, 10 $\mu$m in diameter) embedded in a microparticle swarm is set into rotation by the action of the surrounding inclusions.
 \textbf{b} Glycerol droplets within a predesigned basin of attraction (non-shaded region) are accumulated on a spot. Two droplets aggregated this way are shown in the magnified inset.
 }
\end{figure}

 We can also take advantage of our ability to remotely control the dynamic state of an individual cluster to enable complex mixing patterns in a confined chemical system. As demonstrated for a single aggregate, equally any site in a lattice can be reversibly switched between an arrested aster or a rotating vortex of particles. As example, we show in Fig. \ref{fig:hex_lattice}c (Movie S4) the remote control of just one of the clusters in a square lattice. We arrest the dynamics on this cluster via a spot of UV light, while neighboring clusters remain unaffected. This shows our ability to selectively address a single swarm, and thus to implement arbitrarily complex local mixing patterns. The reported swarming behavior can be employed to actuate on larger embedded objects for which electrophoretic propulsion is not effective. In Fig. \ref{fig:devices}a, a large glass cylinder is set into rotational motion due to the collective interaction with the particles in a vortex, which has been previously nucleated on the cylindrical inclusion. The angular speed of the embedded object can be tuned by varying the speed of the driven particles, which is controlled by means of the applied AC field.

Inclusions of any nature are susceptible to be transported using the strategies described here, with the sole requirement that the NLC director has dipolar symmetry around the colloids. In the case of liquid inclusions, which will spontaneously adopt a spherical shape, the presence of an adsorbed surfactant can be employed to adjust the NLC anchoring and obtain the necessary configuration. In Fig. \ref{fig:devices}b we demonstrate the controlled aggregation of femtosized volumes of glycerol (stabilized with Sodium Dodecylsulphate) into an arbitrary spot. Droplets within a designed basin of attraction can be temporarily stored on a defined location for further processing.

In conclusion, we have reported experiments where colloidal inclusions exhibit reconfigurable emergent behavior and can be assembled and driven in aggregates along arbitrary paths. Ensembles of colloidal particles can be reconfigured into different modes of dynamic self-assembly as they are individually driven in an extended nematic liquid layer. Propulsion is provided by low frequency AC electrophoresis, which is enabled by the anisotropy of the host nematic. The particles move parallel to the local director field, which can be determined over macroscopic distances using simple physico-chemical methods. We control the dynamics of the dispersed particles using reversible photoelastic modulations of the host nematic that can be designed to gather or disperse the colloidal inclusions. Unlike earlier strategies, we can employ the same phoretic mechanism to drive both individual particles or assembled swarms. This ability is crucial for a potential applicability in an confined chemical system, where on-demand transport of precise amounts of chemical cargo to predesigned locations is required. The underlying physicochemical mechanisms rely on generic properties of the involved materials, and can be employed to establish complex local mixing patterns or drive larger embedded objects. Furthermore, since liquid-crystal enabled AC electrophoresis is also a valid transport mechanism for liquid inclusions, we can envisage an extension of the reported strategy to living systems by employing biocompatible lyotropic liquid crystals\cite{zhou2014}. Although biocompatible catalytic micropumps have recently been demonstrated\cite{sengupta2014}, the perspective of control and addressability offered by our strategy are unparalleled.
Finally, and on a broader perspective, the ability to control the dynamic assembly of colloidal inclusions over a surface can lead to novel devices, and can be employed to obtain model systems to study swarming behavior or the dynamics of soft active matter\cite{marchetti2013}.

\clearpage

{\bf EXPERIMENTAL SECTION}\\

Nematic liquid crystal (NLC) cells were prepared using 0.7 mm thick microscope slides of size $15\times25$ mm$^2$, coated with a thin layer of indium-tin oxide (ITO) with a surface resistivity of 100 $\Omega$ per square (VisionTek Systems). The plates were cleaned by sonication in a 1\% Micro-$90$ (Sigma-Aldrich) solution,  rinsed with ultrapure water ($18.2 \,$M$\Omega \cdot$cm, Millipore Milli-Q Gradient A-10), and dried at $110^{\circ}$C for $20 \,$min. To prepare photosensitive plates (Supplementary Section 1), the ITO surface was first coated with a self-assembled monolayer of (3-aminopropyl)triethoxysilane (APTES, Sigma-Aldrich)~\cite{howarter2006}. The surface was subsequently functionalized with the azobenzene 4-octyl-4'-(carboxy-3-propyloxy)azobenzene (8Az3COOH)~\cite{crusats2004}. The synthesis was performed in a dimethylformamide medium (peptide synthesis grade, Scharlau) through an amide group formation between the amino terminal group of APTES and the acid group of 8Az3COOH, using pyBOP ($>97\%$, Fluka) as coupling agent, at a molar ratio $1:1.25$ (8Az3COOH:pyBOP). We measured UV-vis spectra in order to verify the photosensitivity of the resulting azosilane film (Supplementary Fig. S5). To prepare counterplates with homeotropic anchoring, the hydrophilic APTES-coated surfaces were spin-coated with a polyimide compound (0626 from Nissan Chemical Industries, using a 5$\%$ solution in the solvent $26$ also from Nissan) at $2000$ rpm for $10$ s, prebaked 1 min at $80^{\circ}$C to evaporate the solvent and then cured 45 min at $170^{\circ}$C. A photosensitive and a non-photosensitive plate were then separated by a polyethylene terephtalate film (Mylar, Goodfellow, nominal thickness 13 and 23 $\mu$m) and glued together along two sides with the ITO layers facing inwards.

Polystyrene anisometric particles (pear-shaped, $3 \times 4 \, \mu$m, Magsphere) were dispersed in a nematic liquid crystal with negative dielectric anisotropy (MLC-7029, Merck,  room temperature properties: $\Delta \varepsilon \,$(1 kHz)$ \, = \, -3.6$, K$_1$ = 16.1 pN, K$_3$ = 15.0 pN). Cells were filled with freshly prepared dispersions by capillary action and sealed with glue.

Irradiation of the samples at wavelengths $365 \,$nm and $455 \,$nm was performed by means of a custom-built LED epi-illumination setup integrated in an optical microscope. Typical light power densities were $0.1 \,$W/cm$^2$ with a spot size in the sub-millimeter range (Supplementary Section 2). Sinusoidal electric fields were applied using a function generator and an amplifier. Amplitudes were in the range 0 to 35$V$ peak to peak, while frequencies for particle motion were in the range 3 to 50$Hz$.

\clearpage

\bibliographystyle{nature}

\begin{thebibliography}{10}

\bibitem{kim1995}
Kim, E., Xia, Y., and Whitesides, G.~M.
\newblock {\em Nature}{ \bf 376}(6541), 581--584 (1995).

\bibitem{squires2005}
Squires, T. and Quake, S.
\newblock {\em Reviews of Modern Physics}{ \bf 77}(3), 977--1026 (2005).

\bibitem{song2006}
Song, H., Chen, D.~L., and Ismagilov, R.~F.
\newblock {\em Angew Chem Int Ed Engl}{ \bf 45}(44), 7336--56 (2006).

\bibitem{yethiraj2004}
Yethiraj, A., Thijssen, J. H.~J., Wouterse, A., and van Blaaderen, A.
\newblock {\em Advanced Materials}{ \bf 16}(7), 596--600 (2004).

\bibitem{glotzer2007}
Glotzer, S.~C. and Solomon, M.~J.
\newblock {\em Nat. Mater.}{ \bf 6}(8), 557--562 (2007).

\bibitem{jiang2010}
Jiang, S., Chen, Q., Tripathy, M., Luijten, E., Schweizer, K.~S., and Granick,
  S.
\newblock {\em Advanced Materials}{ \bf 22}(10), 1060--1071 (2010).

\bibitem{wang2012}
Wang, Y., Breed, D.~R., Manoharan, V.~N., Feng, L., Hollingsworth, A.~D., Weck,
  M., and Pine, D.~J.
\newblock {\em Nature}{ \bf 491}(7422), 51--55 (2012).

\bibitem{poulin1997}
Poulin, P., Stark, H., Lubensky, T.~C., and Weitz, D.~A.
\newblock {\em Science}{ \bf 275}(5307), 1770--1773 (1997).

\bibitem{whitesides2002}
Whitesides, G.~M. and Grzybowski, B.
\newblock {\em Science}{ \bf 295}(5564), 2418--2421 (2002).

\bibitem{duan2013}
Duan, W., Liu, R., and Sen, A.
\newblock {\em J Am Chem Soc}{ \bf 135}(4), 1280--3 (2013).

\bibitem{zheludev2012}
Zheludev, N.~I. and Kivshar, Y.~S.
\newblock {\em Nat. Mater.}{ \bf 11}(11), 917--924 (2012).

\bibitem{grier2003}
Grier, D.~G.
\newblock {\em Nature}{ \bf 424}(6950), 810--816 (2003).

\bibitem{paxton2004}
Paxton, W.~F., Kistler, K.~C., Olmeda, C.~C., Sen, A., St.~Angelo, S.~K., Cao,
  Y., Mallouk, T.~E., Lammert, P.~E., and Crespi, V.~H.
\newblock {\em Journal of the American Chemical Society}{ \bf 126}(41),
  13424--13431 (2004).

\bibitem{paxton2006}
Paxton, W.~F., Baker, P.~T., Kline, T.~R., Wang, Y., Mallouk, T.~E., and Sen,
  A.
\newblock {\em J Am Chem Soc}{ \bf 128}(46), 14881--8 (2006).

\bibitem{howse2007}
Howse, J.~R., Jones, R. A.~L., Ryan, A.~J., Gough, T., Vafabakhsh, R., and
  Golestanian, R.
\newblock {\em Phys. Rev. Lett.}{ \bf 99}, 048102 Jul  (2007).

\bibitem{wang2009}
Wang, J.
\newblock {\em ACS Nano}{ \bf 3}(1), 4--9 (2009).

\bibitem{buttinoni2013}
Buttinoni, I., Bialk\'e, J., K\"ummel, F., L\"owen, H., Bechinger, C., and
  Speck, T.
\newblock {\em Phys. Rev. Lett.}{ \bf 110}, 238301 (2013).

\bibitem{wang2013}
Wang, W., Chiang, T.~Y., Velegol, D., and Mallouk, T.~E.
\newblock {\em J Am Chem Soc}{ \bf 135}(28), 10557--65 (2013).

\bibitem{gao2014}
Gao, W., Pei, A., Dong, R., and Wang, J.
\newblock {\em J Am Chem Soc}{ \bf 136}(6), 2276--9 (2014).

\bibitem{ibele2009}
Ibele, M., Mallouk, T.~E., and Sen, A.
\newblock {\em Angew Chem Int Ed Engl}{ \bf 48}(18), 3308--12 (2009).

\bibitem{kagan2011}
Kagan, D., Balasubramanian, S., and Wang, J.
\newblock {\em Angew Chem Int Ed Engl}{ \bf 50}(2), 503--6 (2011).

\bibitem{palacci2013}
Palacci, J., Sacanna, S., Steinberg, A.~P., Pine, D.~J., and Chaikin, P.~M.
\newblock {\em Science}{ \bf 339}(6122), 936--40 (2013).

\bibitem{bricard2013}
Bricard, A., Caussin, J.~B., Desreumaux, N., Dauchot, O., and Bartolo, D.
\newblock {\em Nature}{ \bf 503}(7474), 95--98 (2013).

\bibitem{gangwal2008}
Gangwal, S., Cayre, O., Bazant, M., and Velev, O.
\newblock {\em Phys. Rev. Lett.}{ \bf 100}(5), 058302 (2008).

\bibitem{lavrentovich2010}
Lavrentovich, O.~D., Lazo, I., and Pishnyak, O.~P.
\newblock {\em Nature}{ \bf 467}(7318), 947--950 (2010).

\bibitem{lavrentovich2014}
Lavrentovich, O.~D.
\newblock {\em Soft Matter}{ \bf 10}, 1264--1283 (2014).

\bibitem{hernandez2011}
Hern\`andez-Navarro, S., Tierno, P., Ign\'es-Mullol, J., and Sagu\'es, F.
\newblock {\em Soft Matter}{ \bf 9}(33), 7999--8004 (2013).

\bibitem{oswald}
Oswald, P. and Pieranski, P.
\newblock {\em Nematic and cholesteric liquid crystals : concepts and physical
  properties illustrated by experiments}.
\newblock The liquid crystals book series. Taylor \& Francis, Boca Raton,
  (2005).

\bibitem{musevic2006}
Musevic, I., Skarabot, M., Tkalec, U., Ravnik, M., and Zumer, S.
\newblock {\em Science}{ \bf 313}(5789), 954--958 (2006).

\bibitem{wood2011}
Wood, T.~A., Lintuvuori, J.~S., Schofield, A.~B., Marenduzzo, D., and Poon, W.
  C.~K.
\newblock {\em Science}{ \bf 334}(6052), 79--83 (2011).

\bibitem{martinez2011}
Martinez, A., Mireles, H.~C., and Smalyukh, I.
\newblock {\em Proc. Natl. Acad. Sci. USA}{ \bf 108}(52), 20891--20896 (2011).

\bibitem{elastic_constants}


\bibitem{zhou2014}
Zhou, S., Sokolov, A., Lavrentovich, O.~D., and Aranson, I.~S.
\newblock {\em Proc. Natl. Acad. Sci. USA}{ \bf 111}, 1265--1270 (2014).

\bibitem{sengupta2014}
Sengupta, S., Patra, D., Ortiz-Rivera, I., Agrawal, A., Shklyaev, S., Dey,
  K.~K., Cordova-Figueroa, U., Mallouk, T.~E., and Sen, A.
\newblock {\em Nat Chem}{ \bf 6}(5), 415--22 (2014).

\bibitem{marchetti2013}
Marchetti, M.~C., Joanny, J.~F., Ramaswamy, S., Liverpool, T.~B., Prost, J.,
  Rao, M., and Simha, R.~A.
\newblock {\em Reviews of Modern Physics}{ \bf 85}(3), 1143--1189 (2013).

\bibitem{howarter2006}
Howarter, J.~A. and Youngblood, J.~P.
\newblock {\em Langmuir}{ \bf 22}(26), 11142--11147 (2006).

\bibitem{crusats2004}
Crusats, J., Albalat, R., Claret, J., Ign\'es-Mullol, J., and Sagu\'es, F.
\newblock {\em Langmuir}{ \bf 20}(20), 8668--8674 (2004).

\end{thebibliography}

\clearpage

 {\bf Acknowledgements}. We thank Patrick Oswald for the polyimide compound. We acknowledge financial support by MICINN (Project numbers FIS2010-21924C02, FIS2011-15948-E) and by DURSI (Project no. 2009 SGR 1055). S.H.-N. acknowledges the support from the FPU Fellowship (AP2009-0974). P.T. further acknowledges support from the ERC starting grant "DynaMO" (No. 335040) and from the "Ramon y Cajal" program (No. RYC-2011-07605).\\

Supplementary information is available in the online version of this paper.
Correspondence and requests for materials
should be addressed to J. I.-M.~(email: jignes@ub.edu)\\

\clearpage

\end{document}